\documentstyle[preprint,aps,floats,amsfonts,amsmath,epsf]{revtex}

\def\lab#1{\label{#1} }  \def\cit#1{\cite{#1}}

\def\1{\Vec}
\def\e{{\mathrm{e}}}
\def\9{\partial}
\newcommand{\E}{{\mathcal{E}} }
\newcommand{\CO}{{\mathcal O}}
\def\om{\omega}
\def\eps{\epsilon }
\def\lb{\left[ }
\def\rb{\right] }
\def\OM{\Omega}
\def\k{\kappa}

\begin{document}

\draft

\preprint{{\small E}N{\large S}{\Large L}{\large A}P{\small P}-A-544/95,
 hep-ph 9509218}

\title{Spherically symmetric quark-gluon plasma field configurations}

\author{Herbert
Nachbagauer \footnote[3]{e-mail: herby \@ lapphp1.in2p3.fr}}

\address{
Laboratoire de Physique Th\'eorique
ENSLAPP \footnote{URA 14-36 du CNRS, associ\'ee \`a l'E.N.S. de Lyon,
et au L.A.P.P. (IN2P3-CNRS) d'Annecy-le-Vieux}
Chemin de Bellevue, BP 110, \\
F - 74941 Annecy-le-Vieux Cedex,  France}

\date{\today}

\maketitle

\begin{abstract}
We study field configurations in a hot quark-gluon
plasma with spherical symmetry. We show that the electric
fields point into radial direction and solve the effective
non-abelian equations
of motions.  The corresponding charge density  has a localized
contribution which has a gauge invariant interpretation as a pointlike
color charge.  We  discuss  configurations oscillating periodically
in time. Furthermore, we calculate the electric field induced by a
constant local charge that is removed from the plasma for $t>0$
as a model of a decaying heavy quark.

\end{abstract}

\newpage

\section*{Introduction }

Unlike as  QED where linear-response theory supplies a
straight-forward definition of the Debye-screening potential  as the
energy induced by a static delta-like charge distribution, the
non-abelian analog is not a gauge invariant way to couple a classical
charge to the gauge potentials.
Making the perturbed Hamiltonian gauge invariant generally amounts
to include ghost fields which compensate for the trivial
transformation behavior of the classical field strength. It
was argued \cit{Heinz} that particularly in the temporal gauge
($A_0=0$) the gauge potentials and fields are related in a
quasi-abelian way by $\1E =-\9^0\1 A,$ rendering the abelian
response-theory applicable.

The TAG, however, does not admit an
imaginary time formalism (ITF) in the usual way \cit{james}, and
a reformulation in the so-called physical phase space formulation
\cit{nach1} again would call for the inclusion of vertices in the
linear response function necessary to account for the non-abelian
character of the theory.

In order to avoid these difficulties one may consider the electric
dispersion relation without external sources and define the
screening potential as the Fourier-transform of the static propagator
on the physical dispersion.
I want to point out that taking the static limit $\Pi(k_0=0,\1k)$
alone does
not define a gauge invariant potential, and thus a physical
quantity. Although the leading order static self energy is just a
constant, $\Pi_{el}(k_0,\1k)=m^2 =g^2 T^2 (N + N_f/2 )/3$, a next-to
leading order definition requires the inclusion of the physical
dispersion. In the case that the propagator had a pole well separated
from eventually cut contributions, the pole really corresponds to a
physical observable \cit{proof}. However, due to mass-shell
singularities appearing within the standard resummation scheme
\cit{BP}, the longitudinal self energy turns out to be sensible to
a -- perturbatively uncalculable -- magnetic mass \cit{tdes}.
Although the  electric self-energy does develop a gauge invariant
pole at next-to-leading order, it is not clear to what observable
this quantity corresponds.

In this paper we study spherically symmetric field
configurations at leading order which have an interpretation as
generated by a local charge in a gauge invariant way.

\section{Observables for spherical symmetry}

Let us reconsider the generic idea to put a pointlike static quark in
the hot quark-gluon plasma. A classical source may be characterized
by the following properties.\par
\begin{itemize}
\item[(i)]
The source is assumed to have no internal structure such as spin
which would involve a magnetic moment.
\item[\,\,(ii)]
The source is localized and  pointlike, and its location is
static with respect to the rest frame of the plasma.
Since at finite temperature essentially infrared physics is concerned,
one can in fact exclude  the location of the charge  as singular
region from space-time. To make this statement more precise, we
consider physics only at distances greater or equal $T$
away from the hypothetical
classical source.
\end{itemize}

{}From (i) and  (ii) it follows that observables have to be
spherically symmetric. Let us illustrate that those two assumptions are
sufficient in classical electrodynamics to single out the Coulomb
potential as the only solution, up to a multiplicative factor.
The fields satisfy the sourceless Maxwell equations,
\begin{eqnarray}
\1 \9 \1E = 0,\quad \1 \9 \times \1B - \Dot{ \1E}=0,\quad
\1 \9 \times \1E+ \Dot{\1B} =0 ,\quad \1 \9 \1B  =  0
 \lab{max1}
\end{eqnarray}
and are spherically symmetric. Due to the famous theorem saying
that 'you cannot comb a hedgehog',
spherically symmetric vector fields can always be written as a gradient
of a scalar in three dimensions, to wit
\begin{eqnarray}
 \1E(r,t) = \frac{\1r}{r} E(r,t) \quad & \Rightarrow& \quad
 \1E(r,t) = \1\9 f(r,t)  \nonumber  \\ \1B(r,t)  =
\frac{\1r}{r} B(r,t)\quad & \Rightarrow & \quad \1B(r,t)
 = \1\9 g(r,t) . \lab{spher}
\end{eqnarray}
It follows from the Gauss law that
\begin{equation}  \Delta f(r,t)=0 \quad \Rightarrow \quad f(r,t) = a(t) +
\frac{b(t)}{r} \quad\Rightarrow \quad \1E(r,t) = b(t)\; \1 \9 \frac{1}{r}
\quad \end{equation}
and from the second Maxwell-equation in (\ref{max1})
$\Dot{\1E }= \1\9\times  \1B= \1\9 \times \1\9 g =0 ,$ thus $\Dot
b(t)=0$ and the electric field $\1E \propto\1r /r^3 $ corresponds
to the Coulomb potential.

Turning to the non-abelian case, the symmetry properties
(\ref{spher}) cannot be imposed directly on the fields, since
those are not gauge invariant quantities.
According to their transformation behavior we are able to distinguish
two classes of quantities in a gauge theory,
these which transform inhomogeneously
as the gauge potentials,
\begin{equation}  A_{\mu} \to
 U^{-1} A_{\mu} U  +\frac{1}{ig} U^{-1} \9_{\mu} U
\end{equation}
 and those that transform homogeneously
\begin{equation}  \1E \to  U^{-1} \1E U \lab{homog} \end{equation}
as the electric and magnetic fields, the charge and the
spatial current. That quantities will be called pre-gauge invariant
since the trace of products of them  defines observables.
Operators that
have positive and gauge invariant eigenvalues correspond to physical
quantities, and the eigenvalue is the physical observable.
Considering e.g.\ the eigenvalue problem of the
Lie-group  valued electric fields
\begin{equation} \Hat{\1E} (\1x,t) \left\vert E(\1x) \right\rangle =
\1E(\1x,t) \left\vert E(\1x) \right\rangle \lab{eval}
\end{equation}
where the hermitian field operator $\Hat{\1E}(\1x,t)$ is taken in the
Heisenberg-representation,
it is obvious that the eigenvalue is real and remains unchanged under
unitary (gauge) transformations (\ref{homog}).
In that sense all pre-gauge invariant
quantities  potentially correspond to physical observables.
In order to construct a set of equal time observables, however,
we need those operators to commute with each other. As concerns the
Lie algebra $\mathrm{su(N)}$ one may chose a Cartan-Weyl representation
with the basis elements $T^s,\, s=1,\ldots , r,\,
r=\mbox{rank of the algebra}$,
spanning the  abelian Cartan subalgebra and the remaining $n-r$
generators $T^{\pm a}$ where $a$ is an r-dimensional root vector.
A set of commuting variables is given by the diagonal components
$ \1E^{s}  T^s $ (no sum over s) and the hermitian commutators
\begin{equation}  \left[ E^{i,a} T^a , (E^{i,a} T^a)^\dag \right] =
 E^{i,a}  ( E^{i,a} )^* \;\sum_s a(s) T^s \quad \mbox{(no
 sum over $i$ and $a$) }
\end{equation}  where $a(s)$ is some weight depending on the group
under consideration.
Imposing spherical symmetry restricts the real diagonal components
$\1E^s $ analogously to the abelian case (\ref{spher}), whereas
only the modulus of the complex components $\1E^a $ of the roots is
subject to the symmetry,
\begin{equation}  \1E^a  = \frac{\1r}{r} |E^a (r,t)|
\e^{i\chi^a (r,\Omega,t)}  \lab{roots}      .
\end{equation}
This can also be seen directly by considering the
eigenvalues of the eigenvalue problem  (\ref{eval}) where only products
like $E^a E^{-a} = E^a (E^a)^* $ enter.
Moreover, on account of the commutator  $[T^s,T^a] = a(s) T^a $,
 the phase of one additional non-diagonal color component  in
(\ref{roots}) can be absorbed with the help of the local gauge
transformation $U=\exp (i T^s \chi^a / a(s) )$ which leaves the
abelian part $E^s T^s$ invariant.
This shows, that in a spherically symmetric situation, only
$r+1$ ($N$ for SU(N) ) components of pre-gauge invariant variables
can been chosen to depend on the radial coordinate
and time, but the vector component of the variables has
always to point in radial direction as in the abelian case.

One could also consider the more general situation where external
space-time and internal symmetries mix. For
$\mathrm{SU}( 2) $ as
internal symmetry group, this leads to an enlarged number of
solutions of the Maxwell equations \cit{book},
 but is in principle restricted to the case where
internal and external symmetry groups  are locally
isomorphic i.e.\ \ $\mathrm{O(3)} \sim \mathrm{SU(2)}$.

\section{Spherically symmetric field configurations in a hot
quark-gluon plasma}

Let us now turn to the similar situation in a hot quark-gluon plasma.
For field  strengths of order of the temperature, $A \sim T$, and in
the long wavelength limit, $\9 \sim g T$, the Maxwell-equations
 without external source generalize to \cit{blaiz}
\begin{equation}  \left[ D_{\mu},F^{\mu \nu} \right] =
 j_{\mathrm{ind} }^{\nu} \lab{max2}\end{equation}
where $ D^{\mu}=\9^{\mu} + ig \left[ A^{\mu} , \; . \; \right] $ and
$F^{\mu \nu} = \left[ D^\mu , D^\nu \right]/(ig)$.
In that regime the full non-abelian
 character of the theory enters, and one cannot expand the
 covariant derivative in  a series in the coupling constant.
 The induced current
$j_{\mathrm{ind} }^{\mu}$ describes the response of the plasma to the
non-abelian gauge fields  and has to be a covariantly conserved
quantity
\begin{equation}   \lb D_{\mu}, j^{\mu} \rb = 0 \end{equation}
for consistency with the Maxwell equations
(\ref{max2}). It can be expressed by means of an auxiliary field
$W^{\mu}$
\begin{equation}  j^{\mu}_{\mathrm{ind} }(x)  =
m^2 \int \frac{\mathrm{d}\Omega_v}{4 \pi} v^{\mu} W^0(x,v) \lab{jind}
\end{equation}
where $v^{\mu}=(1,\1v) $ denotes a null vector and the
angular integral $\int\mathrm{d}\Omega $ runs over all directions of $\1v$.
 $W^{\mu}$
is a solution to
\begin{equation}  \lb v \cdot D,W^{\mu} \rb = F^{\mu \nu} v_{\nu} \lab{weq}
 \end{equation}
and conjugate to  $v^{\mu},$   $ v_{\mu} W^{\mu}  =0 .$
The set of equations (\ref{max2},\ref{jind}) and (\ref{weq}) forms a
complicated highly non-linear system of integro-differential equations
describing the long wavelength behavior of a hot quark-gluon plasma.
A particular class of plain-wave solutions where the gauge potential
depends on space-time only as a function of $(p\cdot x)$ has been found
in  \cit{blaiz2}.

Here, we will consider spherically symmetric configurations with vanishing
magnetic fields $B^i = \eps^{ijk} F_{jk}=0 $. Note that under a gauge
transformation $\1B$ transforms homogeneously as the electric field
(\ref{homog}) so that solutions with vanishing magnetic fields have a
gauge invariant meaning. Moreover, those configurations are singled out in
the sense that the spatial components of the  corresponding gauge
potentials can always be written as pure gauge,
\begin{equation} \1 A = \frac{1}{ig} V^{-1} \1\9 V , \lab{ahom}
\end{equation}
with $V$ being a space-time dependent element of the gauge group.
In that case the gauge potentials even become pre-gauge invariant
quantities. This class of solutions could also have been obtained by
assuming the gauge fields to be spherically symmetric in the first place,
which also leads to vanishing magnetic fields.
However, as pointed out above, gauge potentials a priori have no
meaning as observables and
thus assuming them to be radially symmetric is rather an ansatz than a
necessary consequence imposed by a symmetry condition.

With the particular form of the potential (\ref{ahom}) it is now
possible  to find radially symmetric solutions of the
Maxwell equations.
For this purpose we note that the covariant derivative of a group
valued  quantity
$X$ can be rewritten in the form
\begin{equation}  \lb \1D ,X \rb = V^{-1} \left( \1\9 ( V X V^{-1} )
 \right)V , \end{equation}
which suggests the introduction of the gauge equivalent quantities
\begin{equation} \1{ \mathcal{E}} = V {\1 E } V^{-1} ,\quad
{\mathcal{Q }} = V \rho_{\mathrm{ind}} V^{-1} ,\quad
{\mathcal{W}}^0 = V W^0 V^{-1}. \end{equation}
Thus the Gauss-law  and the zeroth component of (\ref{weq})
\begin{equation}[ \1D, \1E] = \rho_{\mathrm{ind}} ,
\quad \lb v \cdot D , W^0 \rb = F^{0i} v_i = -\1E \1v
\end{equation}
assume quasi-abelian form
\begin{equation} \1\9  \1{ \mathcal{E}}  =  { \mathcal{Q }} ,\quad
( v \cdot \9 ) \;  {\mathcal{W}}^{0} = -\1{\mathcal{E}} { \1v}.
\lab{chdef}
\end{equation}

We may further choose the spatial components of $v$ to
point into radial direction, $\1v = \1r/r .$
This entails that the spatial components of the induced current
in (\ref{jind}) vanishes and the induced charge becomes proportional to
 ${\mathcal{W}}^0$,
\begin{equation}
{ \mathcal{Q }}(r,t) = m^2 \int \frac{\mathrm{d}\Omega}{4\pi}
{\mathcal{W}}^0(r,t) = m^2 { \mathcal{W}}^0 (r,t) .
 \end{equation}
Due to the linearity of Eq.\ (\ref{chdef}) and the particular form
(\ref{roots}),  the diagonal color
components of the electric field satisfy the
differential equation
\begin{equation}  (\9_0 + \9_r )( \frac{2}{r} + \9_r ){\mathcal{E} }  =
 m^2 {\mathcal{E} } \lab{diffeq}
\end{equation}
which will be in the center of our further investigations.

We construct solutions in terms of eigenfunctions
of the time derivative, ${\mathcal{E}} (r,t)={\mathcal{E}} (r)
\e^{i \omega t}$ where the separation parameter $\om$  may in general
assume complex values. Eq.\ (\ref{diffeq}) can  be straightforwardly
transformed into the radial Schr\"odinger-equation for the hydrogen
atom
\begin{equation}
\frac{d^2}{d\rho^2} \left( \e^{i\om r/2} \E(\rho) \right) +
\frac{2}{\rho} \frac{d}{d\rho}
 \left( \e^{i\om r/2} \E(\rho) \right)+\left( -\frac{2}{\rho^2} +
\frac{i\om}{\k \rho} -\frac{1}{4} \right) \left( \e^{i\om r/2} \E(\rho)
\right)= 0
\end{equation}
with $\k = \sqrt{4 m^2  - \om^2}$ and $\rho=\k r$.
The  solution of (\ref{diffeq}) is composed of a singular and a regular part,
\begin{equation}
{\mathcal{E}} (r) = \frac{\e^{-(\k + i\om) \frac{r}{2}}}{4 \pi r^2}
\left(  \frac{ C_{1}}{2}
 \Gamma ( 2 - i\frac{\om}{\k}) U (-1 - i\frac{\om}{\k} ,-2,\k r )  +
 C_{2}  r^3 M(2-i\frac{\om}{\k},4,\k r ) \right)
\lab{sol}
\end{equation}
where $U$ and $M$ denote the confluent hypergeometric functions
\cit{abragr}, $C_1,C_2$ are constants and the additional factor in
front of the first term  is chosen such that the
series in $r$ of the first summand in the brackets starts with unity.

The quantity $ i\om/(\k \rho ) $ which is the charge of the
nucleus in  the  corresponding quantum mechanical problem, is an
imaginary number for real $\om$. This entails that $\k$ becomes
purely imaginary when the first argument in the hypergeometric function
$M$ is a negative integer and its power series terminates.
Consequently, no stationary regular bounded solutions -- corresponding to the
quantum mechanical bound states -- exist.

The singular part proportional to $U$ which is usually
excluded by the demand of regularity at the origin  has in fact an
interpretation as the electric field generated by a pointlike source.
 Let us 'measure'
the charge corresponding to that field according to (\ref{chdef}).
More precisely, we consider the divergence of
$\E$ in the sense of distributions acting as a functional on
$C_{0}^\infty({{\Bbb R }}^3)$ test functions $\varphi$,
\begin{equation}
\left({ \mathcal{Q}} , \varphi \right)=
- \left(  \vec {\mathcal{E}} ,\vec\9 \varphi  \right) := -\int d\OM
\int_{0}^{\infty} dr r^2 \vec
{\mathcal{E}}(r) \vec \9   \varphi (r\OM)
\end{equation}
which by integration by parts leads to
\begin{equation}
\lim_{r \to 0} (4 \pi r^2 \E(r))\varphi (0) +\int d\OM \int_0^\infty dr r^2
\left( \frac{(r^2 \E(r) )'}{r^2} \right)
\varphi(r\OM)
=: C_1 ( \delta^3,\varphi )  + (\vec\9 \vec\E , \varphi ).
\end{equation}
The resulting charge is  composed of a pointlike source
$\propto  \delta^3 (r) $
which gets contributions only from the singular part of the electric
field and the divergence which corresponds to the charge induced
 by the plasma.
In what follows we will not consider the true vacuum solution
$\propto M$ and concentrate on to physically more interesting situations
with local charges.  Particularly, we focus on an
oscillating point charge and a charge which
is constant for $t < 0$ and vanishes for $t\ge 0$.

\subsection{Monochromatic oscillations}

One particularly interesting class are stationary plasma waves induced
by a periodically oscillating charge,
${\mathcal{Q}}^{\mathrm{local}}(r,t;\om)= \delta^3 (r) \e^{i\om t} $
which electrical field is given by (\ref{sol})
multiplied by $ \e^{i\om t}$ with $C_1 =1$ and $C_2=0$.
Above the frequency $\bar\om = 2 m$,
the inverse damping length $\k$ becomes an
imaginary quantity, and we choose the root such that $\Im (\k) < 0$.
The exponentially decreasing radial
dependence in (\ref{sol}) turns into a radially periodic one. The
hypergeometric function $U$ increases for large radial distances
like $r^{1+i\om /\k } $ which combines with the prefactor to $r^{-1+i\om
/\k} < r^{-1}  $.

For very large frequencies, $\k \to -i \om$,
the hypergeometric function  independently from  $r$ becomes unity,
and the radial dependence of the exponential vanishes. In this case,
the electric fields $\propto \exp (i \om t) /r^2$ cease to propagate
in the plasma.

In the other extreme of a static charge, i.e. for $\om \to 0$,
the electric field $\propto  \e^ {-m r}(1 + m r)/r^2 $  corresponds
to the exponentially screened  Debye potential.

It is instructive to consider the situation of perturbations with
periods corresponding to mass threshold  $\bar\om$.
In that case the
damping length $1/\k$ becomes infinite and the argument in the
 gamma function in (\ref{sol}) diverges. However, the infinities
 are just compensated by the
hypergeometric function and one finds in the limit $\k \to 0$
\begin{equation}
\E^{\om =2 m}(r,t) = m^2 \frac{(2 i x)^{3/2} }{x^2} \e^{2 i y - i x}
 K_3 (2 \sqrt{2 i x} )
\end{equation}
where $x=m r$, $y=m t$, and
 $K_3$ denotes the Neumann function of imaginary argument.
The large distance behavior is somehow in between the static
exponential decrease and the high frequency power-law behavior.
In fact,  asymptotically expanding the Bessel function,
\begin{eqnarray}
\E^{\om =2 m}(r,t) &=& \frac{\sqrt{\pi} m^2 }{2 x^2}
{(2 i x)^{5/4}}\thinspace \e^{ 2 i y - i x- 2 \sqrt{2 i x} } \left( 1 +
 \frac{35}{16\sqrt{2 i x}} + \CO \left( \frac{1}{x} \right) \right)
 \nonumber \\ &=&
\frac{\sqrt{\pi} m^2 }{2 x^2} {(2 i x)^{5/4}} \thinspace
\exp \left( 2 i y - i x- 2 (1+i) \sqrt{ x} - \frac{i\pi}{8}  +
\frac{35  (1-i) }{32
\sqrt{x} } + \CO\left( \frac{1}{x} \right) \right) \end{eqnarray}
shows that the field is damped with an exponential $\propto\sqrt{x}$
only.
The phase gets also  contributions from $\sqrt{x}$,
a constant term and inverse powers of the root.
Asymptotically, constant phases propagate with velocity
$$v_{\phi} = \frac{dx}{dy} = 2 \left(1 -x^{-1/2}  +
x^{-1}  + \frac{29}{64} x^{-3/2 } + \CO( x^{-2})\right) $$
which approaches twice the speed of light for $x\to \infty$.
Physical information about the propagation of waves in the plasma
may be deduced from the acceleration of the phase velocity,
$$ \frac{dv_{\phi}}{dy}=v_{\phi} \frac{dv_{\phi} }{dx }
 \sim 2 x^{-3/2} - 6 x^{-2}+\frac{279}{32} x^{-5/2}   + \CO (x^{-2}). $$
The leading coefficient has a positive sign which means that a
maximum of the oscillation runs slightly faster away from the origin
than its neighboring maximum remoter from the source.
A similar study in the regime $x \lesssim 1$ shows that also  there
waves appear accelerated in positive radial direction with a maximal
acceleration at $x \sim 0.3 $ (Fig1.).

\begin{center}
\begin{figure}
\vspace*{-2em}
\epsfxsize=6.5in  
\epsfbox{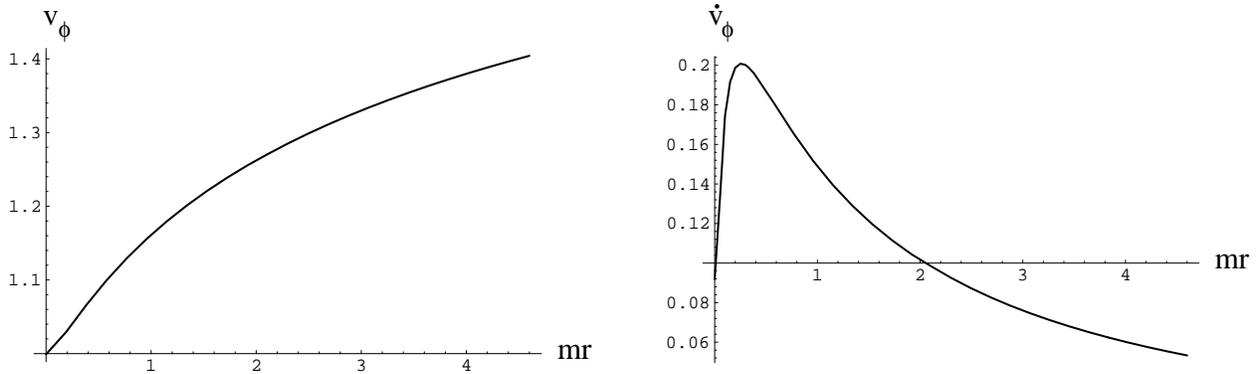}
\vspace*{-2em}
\caption{Phase speed (left) and phase acceleration (right) of the
electric field
in function of the distance of a localized charge
periodically oscillating with frequency $2 m$.}
\end{figure}
\end{center}

\subsection{Shock wave of a suddenly removed point charge}

In this section we study the situation where the localized charge is
constant for $t < 0$ and vanishes for $t\ge 0$. In a physical
experiment, the charge may be a heavy quark which due to its
large mass can be thought of as a well localized classical color source.
 At $t=0$
the quark may decay into a number of lighter particles which can
escape the plasma or be thermalized depending on their energy, the
temperature and the size of the plasma. In any case, we are faced
with a scenario of a -- with respect to the typical time scale $m$ of
the plasma --  suddenly disappearing color-charge.

 To model
the situation we have to find a superposition of the monochromatic
oscillating solutions studied in the last section.
This can be easily achieved by observing that the charge can be
written as
\begin{equation} \delta^3 (r) \theta (-t) = \delta^3 (r)
\frac{i}{2 \pi} \int_{\mathcal C} d\om \frac{\e^{i\om t} }{\om}
=  \frac{i}{2 \pi}\int_{\mathcal C} \frac{d\om}{\om} Q^{\mathrm
local} (r,t;\om).
 \end{equation}
The path of integration is parallel to the real $\om$ axes as shown in
Fig.~2, and  picks up  the residue when closed in the lower half plane.
 The corresponding superposition of the monochromatic
electric fields becomes
\begin{equation}
\E^{\mathrm sw}(r,t)= \frac{i}{16 \pi^2 r^2} \int_{\mathcal C}
d\om \frac{\e^{ i \om t -(\k + i \om ) \frac{r}{2} } }{\om } \Gamma (2
-\frac{i\om}{\k} ) U (-1-\frac{i\om}{\k} ,-2,\k r).
\end{equation}
Prior to evaluating this integral by closing the contours above and
below respectively, we have to define the root $\k = \sqrt{4 m^2 -
\om^2} $ for complex values of $\om$. It admits to branch
points and it turns out to be most useful to chose the cuts parallel
to the imaginary $\om$-axes. The sign is chosen
that $\k \to i\om $
for large modulus of $\om$ and $ |\Re (\om ) | > 2 m$.

\begin{figure}
\epsfxsize=6.5in
\epsfbox{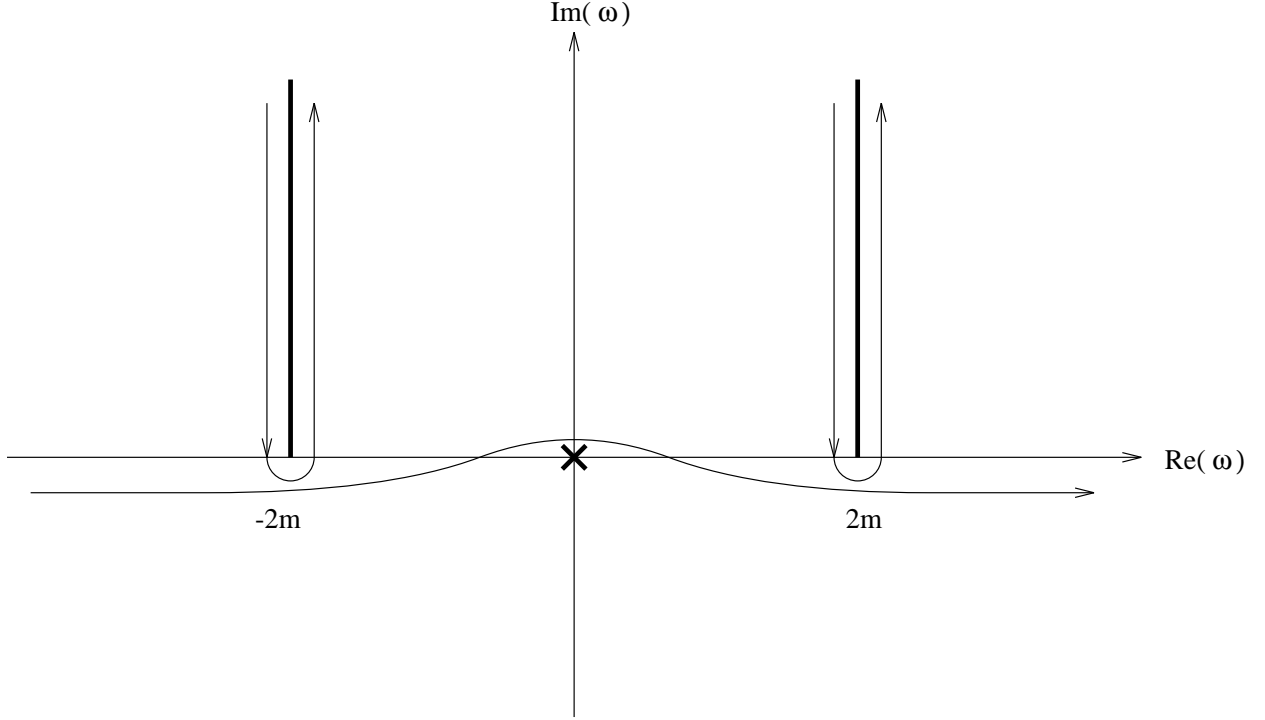}
\caption{Cut and pole structure in the $\omega$-plane of the
decaying-particle solution. }
\end{figure}

This choice admits to close the contour in the lower half plane for
$t-r < 0$ where the integrand vanishes exponentially on the arc. The
only contribution comes from the residuum which gives the
static limit
\begin{equation}
\E^{t<r}(r,t) = \frac{\e^{-m r} }{4 \pi r^2} (1+ mr )
\end{equation}
corresponding to the Debye potential.

For $t>r$ we may deform the contour to the arc $|\om|
\to\infty, \, \Im(\om) > 0$ on which the  integrand vanishes but
which picks up two cut-contributions.
The contours around the cuts are well separated from the pole which
facilitates considerably an asymptotical and numerical
evaluation of the integration. In order to calculate the discontinuity
 we need the relation
\begin{eqnarray}
 \Gamma (2-a)  U (-1-a,-2,z)
& - & \Gamma(2+a) U (-1+a,-2,-z)\e^z \nonumber \\
&=& \frac{a (a^2 -1) \pi}{6 \sin \pi a} z^3 M(2-a,4,  z)
\end{eqnarray}
which can readily be  derived from the integral representation of the
hypergeometric  function.

Coosing the parametrization $\om= 2 m (-1 +i u )$ for the left
contribution, we get
\begin{eqnarray}
\E^{t>r} (r,t) =- m^2  \frac{i x}{12 \pi} \int_0^\infty&&du
 \frac{ \exp (-2 i y + i x -2 y u +  x u + x \sqrt{u}\sqrt{u+2 i} )}
 { \sin\pi \frac{i+u}{\sqrt{u}\sqrt{u+2 i} } } \times
\nonumber \\
&& M(2 - \frac{i+u}{\sqrt{u} \sqrt{u+2 i}
},4,-2 x \sqrt{u}\sqrt{u+2 i} ) + \mathrm{h.c.} \lab{shi}
\end{eqnarray}
where the complex conjugate part stems from the right cut.
This expression is a  particular superposition of regular
solutions corresponding to vanishing local charge since the
causal information of the decay  propagates
in the future directed light cone $t>x$ of the event at $(x=0,t=0)$.

The evaluation of the integral (\ref{shi}) is quite complicated since
the integration variable appears not only in the argument but also in
the index of the hypergeometric function.
For values of $x \sim 1$ and larger, the situation is quite
hopeless since even  the asymptotic expansion of $M$ for large arguments
is spoiled by the diverging index at small $u$ where the integrand
contributes most.

\begin{figure}
\epsfverbosetrue
\centerline{\epsfbox[100 40 352 397]{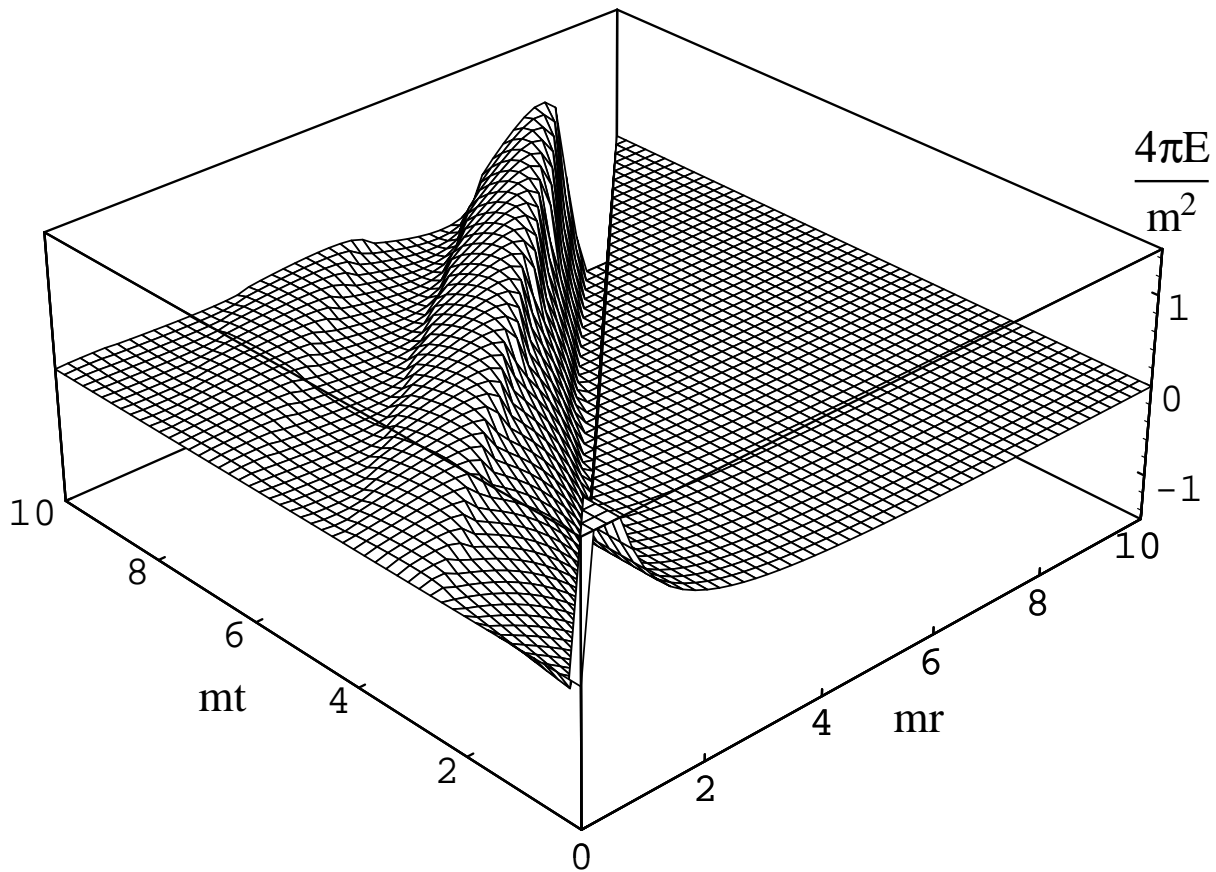} }  
\caption{The electric shock wave propagating in the plasma. Below the
light cone $t<r$ the electric fields correspond to the Debye
potential, are hit by the shock wave at $t=r$ and vanish
asymptotically for late times in the forward-directed light-cone.}
\end{figure}

However, a numerical integration is still possible with the result
shown in Fig.\ 3. The graph shows that waves are damped
for large distances in
direction orthogonal to the light cone. In the region close to the
light cone $r \sim t$ the amplitudes increase with the distance of
the origin.

\newpage

\section{Summary and Conclusion}

We studied the spherically symmetric solutions of electric field
configurations in a hot quark-gluon plasma. We showed assuming vanishing
magnetic fields, which is a necessary compatibility condition with
the symmetry imposed, that the electric fields which are a priori no
physical observables in a non-abelian theory can nevertheless
be given a  physical
interpretation. As in the abelian case, they are radially directed
vector fields. Their physical components can  be gauge
transformed to depend on time and
radius only. Using this property together with the
equations of motions  in the  hard thermal loop approximation,
we construct a differential equation for the fields.
The solution of that equation consists of a singular and a regular
part. Evaluating the corresponding charge in the sense of
generalized functions shows that the singularity at the
origin corresponds to pointlike charge, and the remaining part has an
interpretation a induced charge. This admits to uniquely
relate localized charge, induced charge and corresponding electric
field in a manifestly gauge invariant way.

We focus on two particular interesting physical situations:
A periodically oscillating and a suddenly removed
local charge. In the first case, we establish the
electric field and  concentrate on oscillations with frequencies close to
the frequency $2 m_{\mathrm Debye}$. The
asymptotic field is found to oscillate with amplitude
$ (mr)^{-3/4}\thinspace \e^{ -2 \sqrt{ m r}}$, which is in between the
exponentially damped static limit corresponding to the Debye
potential and a power-law behavior for oscillations above $\om=2 m$.

In the second example we study a suitably chosen superposition of the
periodic solution which has constant local  charge  for $t<0$ and
vanishing  local charge for $t>0$. The electric field is found
to correspond to the static Debye case for $t<r$. In the causal
region $t>r$, the electric field can be written as an
integral which we evaluate numerically.
The suddenly disappearing charge may be regarded as a model for a
heavy, well localized quark placed in a hot quark-gluon plasma which
spontaneously decays into a number of lighter particles which can
escape the plasma  or be thermalized.

We found a gauge invariant concept to replace the  QED-inspired idea
of putting a local classical source into a plasma that potentially
violates
 gauge invariance in the non-abelian case. We find that
symmetry properties, and in particular spherical symmetry, already
single out a class of solutions which admit {\em a posteriori} an
interpretation of a localized charge. That result encourages to look
for a symmetry-based definitions of physical observables, and in
particular of the Debye mass.
Work in that direction is currently under progress.

\section*{Acknowledgements}

This work was supported by the Austrian ``Fonds zur F\"orderung der
wissenschaftlichen Forschung (FWF)'' under project no.\
P10063-PHY, and the EEC Programme "Human Capital and
Mobility", contract CHRX-CT93-0357 (DG 12 COMA).

\end{document}